\documentclass[manuscript]{aastex63}
\usepackage{txfonts}
\usepackage{latexsym,bm}
\usepackage{url}
\usepackage{color}
\usepackage{colortbl}
\usepackage{multirow}
\usepackage{ulem} 
\usepackage{diagbox}
\usepackage{threeparttable}

\newcommand{\kongemail}{kongfanjing@ncwu.edu.cn}
\newcommand{\hit}{School of Science, Harbin Institute of Technology, Shenzhen,
518055, China}
\newcommand{\ncwu}{School of Electronic Engineering, North China 
University of Water Resources and Electric Power, Zhengzhou, 450046, China}

\newcommand{\hitqin}{\hit}
\newcommand{\ncwukong}{\ncwu; \kongemail}
\definecolor{mygray}{gray}{.9}
\shorttitle{electron acceleration efficiency by shocks}

\shortauthors{Qin et al.}

\begin{document}

\title{The efficiency of electron acceleration by ICME-driven shocks} 

\author[0000-0002-3437-3716]{G. Qin}
\affiliation{\hitqin}

\correspondingauthor{F.-J. Kong}
\email{\kongemail}
\author[0000-0001-7617-8268]{F.-J. Kong}
\affiliation{\ncwukong}
\affiliation{\hitqin}

\author[0000-0002-5776-455X]{S.-S. Wu}
\affiliation{\hitqin}

\begin{abstract}
We present a study of the acceleration efficiency of suprathermal electrons at 
collisionless shock waves driven by interplanetary coronal mass ejections
(ICMEs), with the data analysis from both the spacecraft observations and
test-particle simulations. The observations are from the 3DP/EESA instrument
onboard \emph{Wind} during the 74 shock events listed in Yang et al. 2019, ApJ,
and the test-particle simulations are carried out through 315 cases with
different shock parameters. A total of seven energy channels ranging
from 0.428 to 4.161 keV are selected. In the simulations, using a 
backward-in-time method, we calculate the average downstream flux in the
$90^\circ$ pitch angle.
On the other hand, the average downstream and upstream fluxes in the $90^\circ$
pitch angle can also be directly obtained from the 74 observational shock
events. In addition, the variation of the event number ratio with downstream to
upstream flux ratio above a threshold value in terms of the shock angle 
(the angle between the shock normal and upstream magnetic field),
upstream Alfv$\acute{\text e}$n Mach number, and shock compression ratio is
statistically obtained. It is shown from both the observations and simulations
that a large shock angle, upstream Alfv$\acute{\text e}$n Mach number, and shock
compression ratio can enhance the shock acceleration efficiency. 
Our results suggest that shock drift acceleration is more efficient in the 
electron acceleration by ICME-driven shocks, which confirms the findings of 
Yang et al. 2018.

\end{abstract}


\section{INTRODUCTION}

Electron acceleration is a vital topic in space and astrophysical plasmas.
The well-known efficient accelerators are collisionless shocks which are
able to accelerate electrons to high energies. It is widely accepted that
diffusive shock acceleration
\citep[DSA;][]{Axford1977,Krymsky1977,Bell1978,Blandford1978}, i.e., the
combination of first-order Fermi acceleration (FFA) and shock drift
acceleration (SDA), remains the dominant particle acceleration 
mechanism at collisionless shocks. Recently many authors have
studied the acceleration of electrons. 
\citet{Tsuneta1998} proposed that in solar flares the formation of an 
oblique fast shock below the reconnection region provides the acceleration
site for nonthermal electrons of 20--100 keV by first-order Fermi acceleration
(FFA) mechanism. \citet{Guo2012} studied electron acceleration at the flare
termination shock in large-scale magnetic fluctuations and found that 
electrons can be accelerated to a few MeV which can qualitatively explain the
observed hard X-ray emissions. \citet{Mann2001} suggested that in the solar
corona, high energetic electrons of 1 MeV can be produced by shock waves 
if strong magnetic field fluctuations appear near the shock transition
so that electrons are allowed to cross the shock front many times by field-line
meandering. 
\citet{Klassen2002} investigated the origin of 0.25--0.7 MeV electrons in
solar energetic particle events through observations from COSTEP/\emph{SOHO} 
and \emph{Wind} 3DP instruments and concluded that the electrons measured
are accelerated by coronal shock waves.
\citet{Mann2006} proposed that hard X- and $\gamma$-rays can be generated 
by electrons that are accelerated by shock drift acceleration (SDA) mechanism
at the termination shock during solar flares.
\citet{Warmuth2009} developed a quantitative model based on radio and 
hard X-ray observations to study the acceleration of electrons in solar flares, 
and showed the possibility of shock drift acceleration at the reconnection 
outflow termination shock. In addition, \citet{Miteva2007} proposed that 
in the solar corona energetic electrons at quasi-perpendicular shocks can be 
accelerated by resonant whistler wave-electron interaction.
\citet{Saito2011} performed particle-in-cell simulations for electron 
acceleration at a quasi-perpendicular shock, and showed that the parallel 
scattering by the kinetic Alfv$\acute{\text e}$n turbulence, which is generated
in the shock transition, suppresses the reflection of electrons during the 
shock drift acceleration process. 

\citet{Dresing2016} determined the particle acceleration efficiency 
at interplanetary shock crossings observed by \emph{STEREO} spacecraft, 
and found that only $1\%$ (five events) of all analyzed shocks shows 
a shock-associated electron intensity increase for 65--75 keV electrons. 
Among the five events, they found that four were associated with ICME-driven
quasi-perpendicular shocks and the rest one was associated with an SIR 
\footnote{stream interaction region}(with an ICME embedded in) 
forward quasi-parallel shock. It is possible that the observed 
shock-associated electron intensity increase events have more occurrence 
frequency in quasi-perpendicular shocks than in quasi-parallel ones, 
although there is very poor statistics in their studies. Furthermore,
\citet{Yang2018} studied the acceleration of suprathermal electrons in the 
energy range of 0.3--40 keV at an ICME-driven quasi-perpendicular shock with
the spacecraft observations, and obtained the $90^\circ$ pitch angle
enhancements and a larger downstream electron spectral index. In the statistical
work of \citet{Yang2019}, for those selected quasi-perpendicular and
quasi-parallel shock events they found a positive correlation between the
average downstream suprathermal electron flux and magnetosonic Mach number,
and the downstream flux enhancement and shock compression ratio. These
results implys the importance of shock drift acceleration in the acceleration
of electrons at quasi-perpendicular shocks. In addition, \citet{Yang2019} 
concluded that there are more SEP events associated with quasi-perpendicular
shocks compared to quasi-parallel shocks.

In the previous work \citep{Kong2020}, using test-particle simulations, we
investigated the acceleration of suprathermal electrons at a 
quasi-perpendicular shock studied by \citet{Yang2018}, suggesting the 
importance of shock drift acceleration in the acceleration of electrons. In
this paper, using spacecraft data analysis and test particle simulations, 
we expand to study statistically 
the effect of shock parameters, such as shock angle (the angle 
between the shock normal and upstream magnetic field), upstream
Alfv$\acute{\text e}$n Mach number, and shock compression ratio, on the
ICME-driven shock acceleration efficiency. In addition, the theoretical models 
for shock acceleration efficiency are derived to compare with the observations and
simulations. In Section
\ref{sec:obs}, we show the observational data. We describe the numerical model
in Section \ref{sec:nmodel}, and the models for efficient shock acceleration in
Section \ref{sec:theory}. The results of observational analysis and simulations
are presented in Section \ref{sec:results}. Summary and discussion are given in
Section \ref{sec:summary}.

\section{Observational DATA}
\label{sec:obs}

In the survey of \citet{Yang2018}, 74 ICME-driven shocks are 
observed by \emph{Wind} over the period from 1995 to 2004. 
The electron electrostatic analyzer (EESA) in the 3DP instrument
\citep{Lin1995} onboard \emph{Wind} 
provides three-dimensional (3D) data for eight pitch angle channels in the 
energy range of $\sim$3 eV to $\sim$30 keV. For the shock events, the flux 
data of energetic electrons in the upstream and downstream of the shock 
(i.e., before and after the shock arrival) can be obtained from the
\emph{Wind}/3DP/EESA instrument. 

\begin{figure}
\epsscale{.8}
\plotone{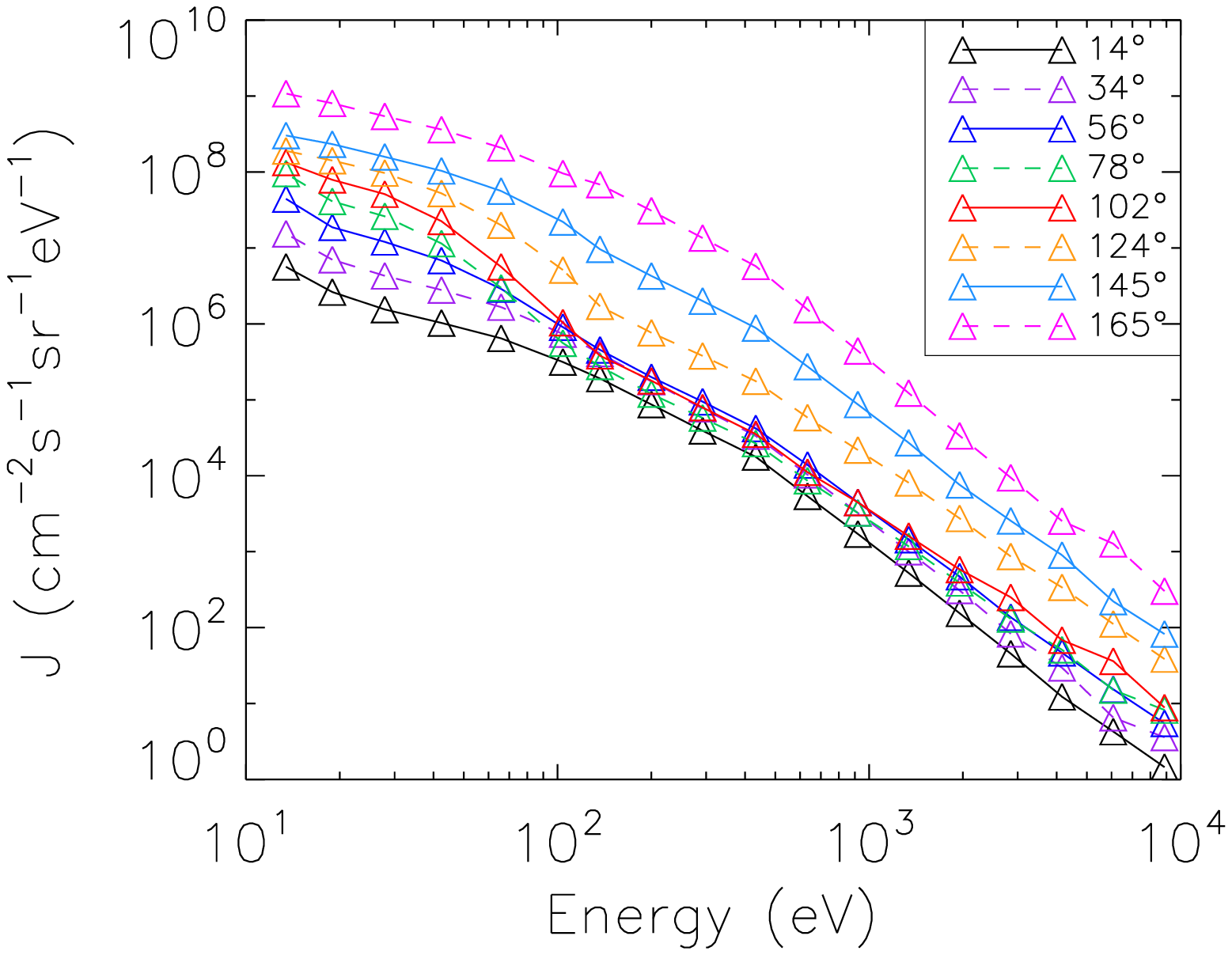}
        \figcaption{Initial upstream electron energy spectra 
for eight pitch angle channels denoted with different color triangles 
by averaging the flux data on \emph{Wind}/3DP ahead of the shock in 
a period of 10 minutes. Note that the spectra with pitch angles of 
$14^\circ$, $34^\circ$, $56^\circ$, $78^\circ$, $102^\circ$, $124^\circ$,
$145^\circ$, and $165^\circ$ are multiplied by $2^0$, $2^1$, $2^2$, $2^3$,
$2^4$, $2^5$, $2^6$, and $2^8$, respectively.
\label{fig:initflux3dp}}
\end{figure}

We choose the shock event on 2000 Feb 11 observed by \emph{Wind}
as an example. The upstream energy spectra in all pitch angle directions 
averaged in a period of 10 minutes (23:14 UT--23:24 UT) prior to the shock are
plotted in Figure \ref{fig:initflux3dp}, multiplied by $2^0$, $2^1$,
$2^2$, $2^3$, $2^4$, $2^5$, $2^6$, and $2^8$ for the pitch angles of
$14^\circ$, $34^\circ$, $56^\circ$, $78^\circ$, $102^\circ$, $124^\circ$,
$145^\circ$, and $165^\circ$, respectively. These spectra are assumed to be
used as the initial upstream electron distribution before the shock
acceleration in the simulations. In this work, we focus on the observations of
the electron energy channels at $\sim$0.428, 0.634, 0.920, 1.339, 1.952, 2.849,
and 4.161 keV indicated by $E_k$ with $k=1$, 2, 3, \dots, 7, as listed in Table
\ref{energychannels}, from the \emph{Wind}/3DP/EESA measurement. Using the data
observed by Magnetic Field Investigation (MFI) \citep{Farrell1995,Kepko1996} 
and Solar Wind Experiment (SWE) \citep{Ogilvie1995} instruments  onboard
\emph{Wind} we obtain the upstream average magnetic field $B_{01}=7.0$ nT and
proton number density $n_{\text{p}}=5.19~\text{cm}^{-3}$. Thus, the upstream
Alfv$\acute{\text e}$n speed is $V_{\text A1}=67~\text{km}~\text s^{-1}$ with 
the formula $V_{\text A1}=B_{01}/\sqrt{\mu_0n_{\text p}m_{\text p}}$,
where $\mu_0$ is space permeability and $m_{\text p}$ is proton mass.

\section{numerical model}
\label{sec:nmodel}

\begin{figure}
\epsscale{1.}
\plotone{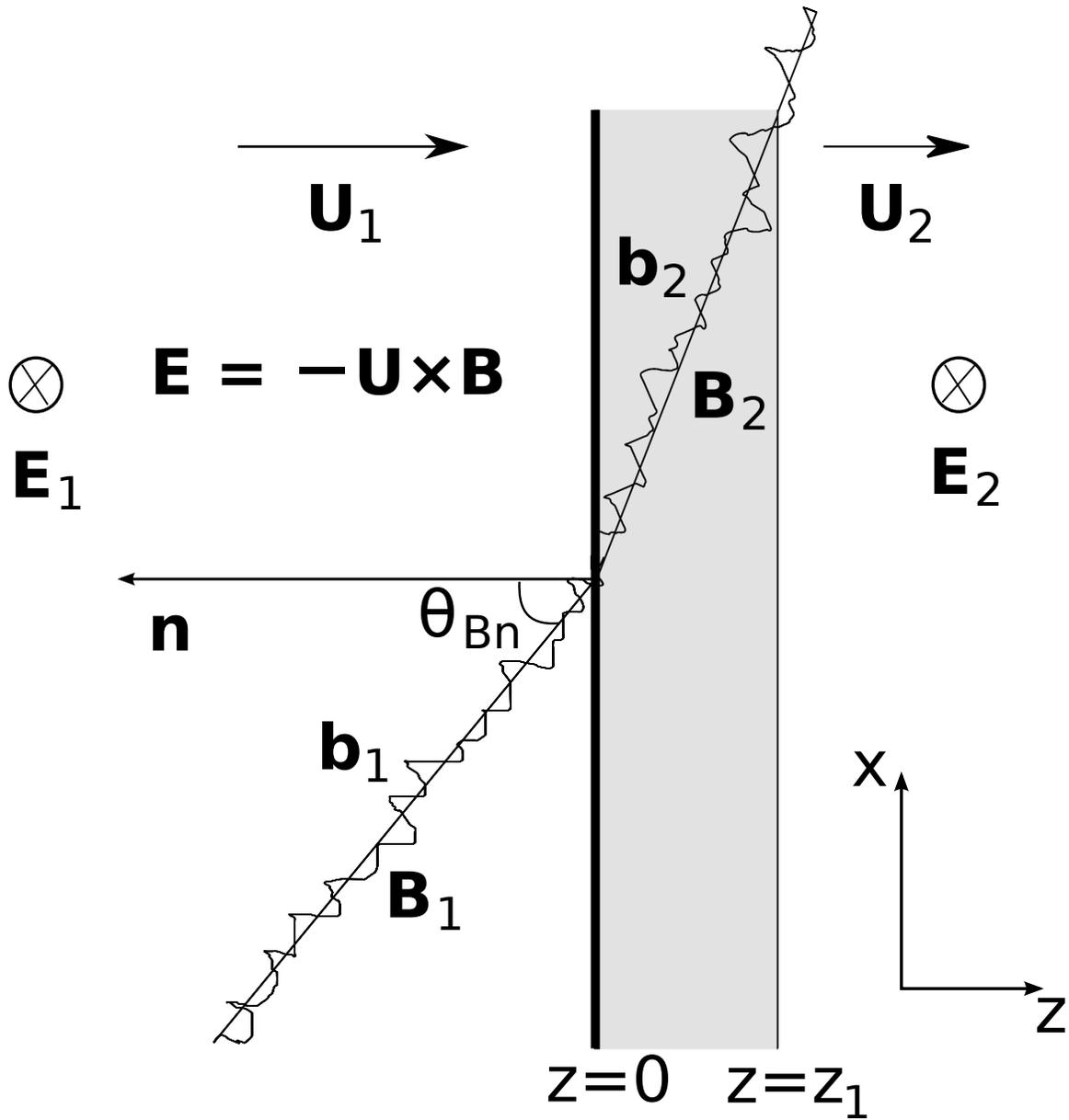}
        \figcaption{The schematic diagram for the shock geometry. The shock
plane is located at $z=0$, and the downstream fluxes from observations and simulations 
are measured in the range of $z$ from 0 to $z_1$. Note that the shock thickness
is not shown.
\label{fig:shock}}
\end{figure}

We study the acceleration of electrons at a 
plane shock by using spacecraft data analysis and test-particle
simulations similar to our previous work \citep{Kong2017, Kong2019}. 
Under the given shock parameters, the trajectories of test particles are
traced by calculating the motion of equation of particles in the
electromagnetic field
\begin{equation}
\frac{d\boldsymbol p}{dt}=q[\bm{E}(\bm{r},t)+\bm{v}\times\bm B(\bm{r},t)],    
\end{equation}
where $\bm p$ is the particle momentum, $q$ is the electron charge, $\bm v$ is
the particle velocity, and t is time. The electric field $\bm E$ is the
convection electric field $\bm{E}=-\bm{U}\times\bm{B}$.
The shock is located at $z=0$, and the plasma flows from the
upstream region with a speed $\bm U_1$ to downstream region with a speed
$\bm U_2$ in the shock reference frame, as shown in Figure \ref{fig:shock} 
(similar to Figure 1 in \cite{Kong2020}).
Here, the upstream speed $U_1$ and downstream speed $U_2$ can be determined by
\begin{equation}
    U_1=M_{\text{A1}}V_{\text{A1}},
    \label{equ:U_1}
\end{equation}
and
\begin{equation}
    {U_2}=\frac{U_1}{s},
    \label{equ:U2U1}
\end{equation}
with the upstream Alfv$\acute{\text e}$n Mach 
number $M_{\text{A1}}$ and shock compression ratio $s$. 
It will be later shown that the downstream
fluxes from observations and simulations are measured in the range of $z$ from
0 to $z_1=2.7\times 10^{-3}$ au. We assume the plasma speed in the shock
transition is in the form \citep[e.g.,][]{Qin2018}
\begin{equation}
U(z)=\frac{U_1}{2s}\left\{\left(s+1\right)+\left(s-1\right)\tanh
\left[\tan\left(-\frac{\pi z}{L_{\text{th}}}\right)\right]\right\},
\end{equation}
where $L_{\text{th}}$ is the shock thickness and assumed to be
$L_{\text{th}}=2\times10^{-6}$ au in this work. It is noted that 
$L_{\text{th}}$ is too small to be shown in Figure \ref{fig:shock}.
The magnetic field, $\bm B$, is given by
\begin{equation}
    \bm{B}(x,y,z)=\bm{B_0}(z)+\bm{b}(x,y,z),
\end{equation}
where $\bm{B_0}$ is a background magnetic field assumed to lie in $x$--$z$ plane,
and $\bm b$ is a turbulent field perpendicular to the background field. 
Note that the magnetic field $\bm B$ is taken to be a 
static magnetic field \textbf{without the flow-advected effect for 
simplicity considering the high speed of particles}
\citep[e.g.,][]{Fraschetti2015}.
The turbulent field, composed of a slab and two-dimensional (2D) components
\citep{Matthaeus1990,Mace2000,Qin2002a,Qin2002b}, is given by
\begin{equation}
    \bm{b}(x',y',z')=\bm {b_{\text{slab}}}(z')+\bm{b_{\text{2D}}}(x',y'),
\end{equation}
where $z'$-axis in the Cartesian coordinate ($x',y',z'$) system, which is 
used to generate the slab and 2D turbulence, is parallel to the background
magnetic field $\bm B_0$.

In all simulations, the slab turbulence has a correlation length of
$\lambda=0.02$ au. The ratio of slab to 2D correlation length is 2.6 according 
to previous studies \citep{Osman2007,Weygand2009,Weygand2011,Dosch2013}. We
define the turbulence in a periodic box of size [10$\lambda$,10$\lambda$] for 
the 2D component turbulence and size 25$\lambda$ for the slab component
turbulence. A dissipation range in which low-energy electrons resonate for the 
slab turbulence is required, and the break wavenumber is fixed at $k_{\text
b}=10^{-6}~\text m^{-1}$ from the slab inertial to dissipation ranges. The
spectral indices of the inertial and dissipation ranges are 
$\beta_{\text i}=5/3$ and $\beta_{\text d}=2.7$, respectively. Note that
the pitch angle scattering corresponds to the parallel diffusion, and the
resonance between particles and turbulence is not affected by the 2D
turbulence assuming weak nonlinear effects. Hence, the dissipation range of 2D
turbulence is ignored. We take the turbulence level, $(b/B_0)^2$, of 0.25 in 
the upstream and 0.36 in the downstream of the shock. 
Note that the turbulent magnetic fluctuations $\bm{b}$ is supposed 
to be perpendicular to the background magnetic field $\bm{B}_0$ and have a 
zero-average in both upstream and downstream. We assume that the enhancement
of the turbulence level, $(b/B_0)^2$, from upstream to downstream, is caused by
the strong disturbance of the shock wave. The detail of the varying
turbulence level throughout the
shock structure is very complicated. \textbf{If the upstream turbulence level is
set, the value of downstream turbulence level has to be obtained with
theoretical modeling depending on values of different parameters, e.g., the
density compression.} However, for the simplicity, we use a
fixed turbulence level in this work to study the acceleration
efficiency of electrons by the interplanetary shock.
In addition, the energy
density ratio of the slab to 2D components is assumed to
$E_{\text{slab}}:E_{\text{2D}}=20:80$. 

In the simulations of shock acceleration of energetic electrons, we
use a method of backward tracing electron trajectories \citep[i.e.,
backward-in-time method,][]{Kong2017,Kong2020}. 
For each simulation case, we want to obtain the downstream distribution
of accelerated electrons under a given initial distribution
(e.g., upstream flux taken as the initial distribution in this work). 
Therefore, 
we put a large number of test particles in the downstream and
calculate the trajectories of those particles backward. After a
period of time, the particles move back to the upstream at the initial time 
$t_0$.
The downstream distribution would be obtained with Equation 5 in 
\citet{Kong2020}
by calculating all the test particles.
In this work each of the simulation cases does not correspond to any of
the 74 shock events from the observations. It
is complicated to obtain all the physical parameters for simulations
in each of the observational shock events. In addition, the initial
pitch angle distributions of all the observed  events are difficult to be
obtained; in particular, several events are even lack of observational data in
some pitch angle directions. Therefore, we do not simulate the 74 observed 
shock events. Instead, we construct a parameter space with the shock angle
$\theta_{\text{Bn}}$, upstream Alfv$\acute{\text e}$n Mach number $M_{\text A1}$, 
and shock compression ratio $s$ (listed in Table \ref{casespara}). This allows us 
to obtain a total of $9\times 7\times 5=315$ virtual shock event cases, for 
each of which 
we perform shock acceleration simulations of electrons. Then, the acceleration 
efficiency of electrons at the shock with different parameters is statistically 
analyzed, and compared with the result from the observations.

At the end of the simulation time $t_{\text{acc}}=10$ min, 
the downstream flux in the range $[z_0,z_1]$ in the $90^\circ$ pitch angle 
for a target energy channel is obtained based on the initial upstream
distributions as shown in Figure \ref{fig:initflux3dp}, where $z_0= 
L_{\text{th}}/2$ and $z_1=2.7\times10^{-3}$ au. 
Note that the value of $z_1$ is set to a distance that a shock with a
speed of 682 km s$^{-1}$ travels within 10 min. It is also noted that the 
speed of 682 km s$^{-1}$ happens to be equal to that of the shock on 2000 Feb
11.  Actually, the exact values of $z_0$ and 
$z_1$ do not change the general findings in this work. More details about the
backward-in-time method are given in Section 3 in \citet{Kong2020}. 
In order to calculate the trajectory of each test particle, we use a fourth
order Runge-Kutta scheme with adaptive time stepping.  Our calculation is
regulated by a fifth order error estimate step with the error parameter 
$10^{-9}$. If we have a higher number of test particles and higher accuracy of
the trajectories of test particles, we can get a more accurate downstream flux from
the simulations. In our previous work \citep{Kong2020}, the downstream flux from
the simulations are compared with the observed downstream flux. While in this
work the simulated shock cases with the artificial shock parameters 
do not correspond to any of the observational events. Therefore, we do not 
directly perform comparisons of downstream flux between the simulations and
observations.
The input parameters for the shock and turbulence in the simulations are 
summarized in Table \ref{shockturbpara}. The turbulence level of 
observations usually becomes greater near the shock surface \citep[e.g.,][]{Kong2017},
and it is not the same in different shock events.
Note that the turbulence level $(b/B_0)^2$ in Table \ref{shockturbpara} 
is set to a fixed value upstream and downstream of the shock
for a qualitative analysis.   
Besides, the ratio of shock thickness $L_{\text{th}}$ to the
gyroradius of the electrons of minimal and maximal energy in the simulations
are 30 and 10, respectively, which implies a small gyroradius of electrons 
compared with the shock thickness, so that the gyro-cycles of energetic electrons 
we study could stay in the shock transition region for a period of time.

\section{Models for efficient shock acceleration}
\label{sec:theory}

Next, we show some qualitative models for the efficient shock
acceleration.
Suppose that particles are accelerated by shock drift acceleration, the
drift time $T_{\text{drift}}$ can be written as  \citep{Kong2020}
\begin{equation}
    T_{\text{drift}}= \frac{L_{\text{th}}}{2U_1}+
        \frac{L_{\text{th}}}{2U_2},
        \label{eq:Tdrift_th}
\end{equation}
here we assume electrons are in a completely scatter-free regime.
\textbf{Note that Equation (\ref{eq:Tdrift_th}) is strictly valid only in the
shock with a perpendicular geometry.}
Furthermore, in order to achieve efficient acceleration, particles should not
move away from the shock transition within the drift time, i.e.,
\begin{equation}
    \frac{1}{2}T_{\text{drift}}v\cos\theta_{\text{Bn}}\lesssim L_{\text{th}},
    \label{eq:cos_theta}
\end{equation}
where the factor $1/2$ in the left-hand side is used to consider the
average over the pitch angle of particle velocity. Note that in
the left hand side of Equation (\ref{eq:cos_theta}) a change of the angle between
the magnetic field and the shock normal across the shock within 
$L_{\text{th}}$ should be taken into account. However, we ignore the angle change 
for simplicity, because it is very difficult to consider this effect. Thus, for the
critical value $\cos\theta_{\text{Bn,c}}$ with efficient acceleration, we have
\begin{equation}
    \cos\theta_{\text{Bn,c}}=\frac{2L_{\text{th}}}{T_{\text{drift}}v}.  
    \label{equ:theta_b}
\end{equation}

On the other hand, \citet{Drury1983} showed that for each cycle of the particle
crossing of the shock front the average momentum change of particles is
\begin{eqnarray}
   \langle\Delta p\rangle&=&2p\int_0^1\frac{\mu(U_1-U_2)}{v}2\mu d\mu\nonumber\\
	{}&=&\frac{4}{3}\frac{U_1-U_2}{v}p.\label{equ:deltap}
\end{eqnarray}
The work done by electric field force is equal to the energy change, i.e.,
\begin{equation}
    w=\frac{p\langle\Delta p\rangle}{\gamma m_0}.
       \label{equ:work2tmp}
\end{equation}
Substituting Equation (\ref{equ:deltap}) into Equation (\ref{equ:work2tmp}),
and considering Equations (\ref{equ:U_1}) and (\ref{equ:U2U1}), we get
\begin{equation}
    w=\frac{4}{3}\left(1-\frac{1}{s}\right)pM_{\text A1}V_{\text A1}.
    \label{equ:work}
\end{equation}
For the qualitative analysis, we suppose that the particle guiding center 
displacement in the shock normal direction during the shock drift acceleration is
$L_{\text {th}}/2$ only considering upstream of the shock.
Then, the number of particle crossings of the shock front is
\begin{equation}
    N_{\text {cr}}=\frac{L_{\text{th}}}{2U_1 T_{\text {gyro,1}}},
        \label{eq:N_cr}
\end{equation}
where $T_{\text{gyro}}$ is the gyro-period of particles.
In order for energetic particles to be efficiently accelerated, the total energy 
increase is assumed to be large enough compared to the particle kinetic energy
$E_{\text{k}}$, i.e.,
\begin{equation}
    N_{\text {cr}}w\gtrsim E_{\text k}.\label{equ:workgg}
\end{equation}
Substituting Equation (\ref{equ:work}) into Equation
(\ref{equ:workgg}), 
we obtain
\begin{equation}
    M_{\text{A1}}\gtrsim\frac{3E_k s}
    {4N_{\text {cr}}pV_{\text {A1}}(s-1)},
    \label{equ:MA1s_M}
\end{equation}
or
\begin{equation}
    s\gtrsim\left(1-\frac{3E_k}{4N_{\text{cr}} pV_{\text {A1}} M_{\text{A1}}}
    \right)^{-1}.
    \label{equ:MA1s_s}
\end{equation}
Assuming the compression ratio $s$ is given by a representative value $s_{\text r}$,
from Equation (\ref{equ:MA1s_M}) we have
the critical value of the upstream Alfv$\acute{\text e}$n Mach number
with efficient acceleration as
\begin{equation}
    M_{\text{A1,c}}=\frac{3E_k s_{\text r}}
    {4N_{\text {cr}}pV_{\text {A1}}(s_{\text r}-1)}.
    \label{equ:M_A1b}
\end{equation}
Similarly, assuming the upstream Alfv$\acute{\text e}$n Mach number $M_{\text{A1}}$
is given by a representative value $M_{\text r}$, from Equation
(\ref{equ:MA1s_s}) we have the critical value of the compression ratio
with efficient acceleration as
\begin{equation}
    s_{\text c}=\left(1-\frac{3E_k}{4N_{\text {cr}}p
    V_{\text {A1}}M_{\text r}}\right)^{-1}.
    \label{equ:s_b}
\end{equation}


Here, we obtain some qualitative models for efficient shock
acceleration for critical values by assuming shock drift acceleration and
quai-perpendicular shocks. 
It is suggested that particles are efficiently accelerated in 
quasi-perpendicular shocks, so these models for critical values can be used for
the qualitative analysis purpose.
\section{results of observational analysis and simulations}
\label{sec:results}

\subsection{Stream of energetic particles anti-sunward}

\begin{figure}
\epsscale{.8}
\plotone{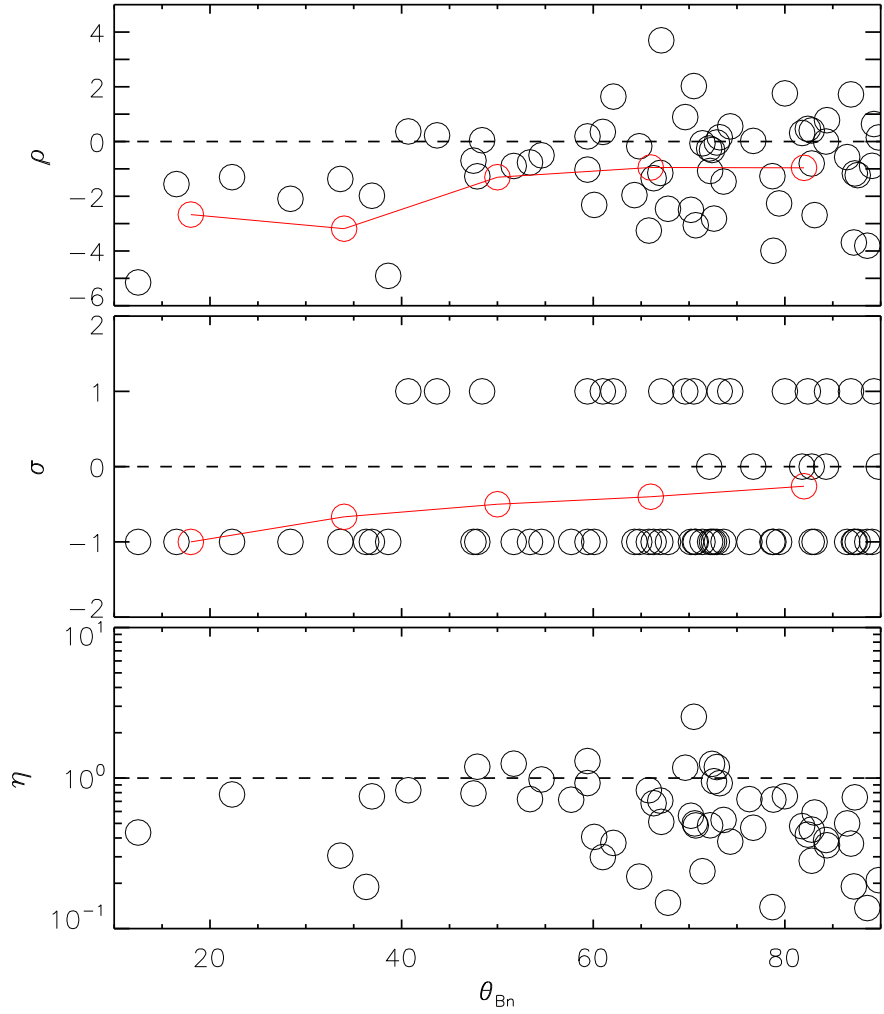}
\figcaption{Top, middle, and bottom panels show parameters $\rho$, $\sigma$,
and $\eta$,
respectively, versus the shock angle $\theta_{\text{Bn}}$ for 65 shock
events with black circles. The red circles in the top and middle panels indicate the
average $\rho$ and $\sigma$, respectively, in each shock angle
interval. \label{fig:rhosigma}}
\end{figure}

According to the previous work by \citet{Yang2018}, it is noted that in 
some SEP events there is a stream of energetic particles in the direction
away from the Sun downstream of the shock  \citep[see also,][]{Kong2020}. 
Here, we check whether this phenomenon is common. For all the 74 
ICME-driven shock events from the data observed by Magnetic Field Investigation
(MFI) onboard \emph{Wind} we determine the magnetic field direction 
toward or away from the Sun by averaging the magnetic field x-component
in GSE coordinates over 10 minutes before and after the shock arrival,
since the GSE $x$ axis points to the Sun. Next, in terms of the pitch angle
$\theta$ we define a modified pitch angle $\theta^\prime$
\begin{equation}
\theta^\prime = \left\{ \begin{array}{ll}
\theta & \rm{if\ magnetic\ field\ is\ sunward},\\
	\pi-\theta & {\rm{otherwise.}} \\
\end{array} \right.\label{thetaprime}
\end{equation}
Therefore, the electrons with a pitch angle $\theta^\prime\sim0^\circ$
are always sunward-streaming regardless of the sunward or anti-sunward direction of 
the interplanetary magnetic field. Using the data observed by the
\emph{Wind}/3DP/EESA measurement, we get the integrated differential intensity
$I_{\theta^\prime}$ over the energy range of 0.428 -- 4.161 keV with 10-minute
average downstream of the shock for each of the modified pitch angle interval
$\theta^\prime$. For the 65 out of 74 ICME-driven shock events from
\citet{Yang2019}, we obtain $I_{0}$, $I_{90}$, and $I_{180}$ with the modified
pitch angle $\theta^\prime$ nearest to $0^\circ$, $90^\circ$, and $180^\circ$,
respectively, and we set $I_{\text{M}}$ as the maximum of $I_0$, $I_{90}$, and
$I_{180}$. Similarly, we also get the integrated differential intensity 
for upstream $I_{\theta^\prime}^{\text{u}}$ with 10-minute average upstream of
the shock. Note that 9 of the 74 shock events are excluded due to the unavailable data 
in some energy channels. Then, we define three parameters $\rho$, $\sigma$,
and $\eta$ as
\begin{equation}
    \rho=\frac{I_{0}}{I_{90}}-\frac{I_{180}}{I_{90}},
\end{equation}
\begin{equation}
\sigma = \left\{ \begin{array}{ll}
1 & \text{if $I_0=I_{\text{M}}$,}\\
0 & \text{if $I_{90}=I_{\text{M}}$,}\\
-1& \text{otherwise,} \\
\end{array} \right.\label{sigma}
\end{equation}
and 
\begin{equation}
\eta=\frac{I_{0}/I_{0}^{\text{u}}}{I_{90}/I_{90}^{\text{u}}}.\label{eta}
\end{equation}
Thus $\sigma=1$ or $-1$ when the electron 
beams are sunward or anti-sunward, respectively.
The top, middle, and bottom panels of Figure \ref{fig:rhosigma} show parameters
$\rho$, $\sigma$, and $\eta$, respectively, versus the shock angle
$\theta_{\text{Bn}}$ for the 65 ICME-driven shock events with black circles. 
The red circles in the top and middle panels indicate the average $\rho$ and $\sigma$, 
respectively, in each shock angle interval. 

In the top and middle panels of Figure \ref{fig:rhosigma} we can see that for most
of the ICME-driven shock events there is a stream of energetic particles away
from the Sun downstream of the shock. For comparison of the simulations and
observations, the stream of particles in the observations has to be avoided. In order 
to study energetic particles accelerated by the shock not in the perpendicular
direction, we need to choose the sunward particles. In the bottom panel of Figure
\ref{fig:rhosigma}, it is seen that for most observational cases the acceleration
of electrons is much less efficient in the sunward direction than in the
perpendicular direction. Therefore, we suggest that the strongest acceleration
occurs in the perpendicular pitch-angle direction, which is consistent with 
the results from both the observations \citep{Yang2018} and simulations \citep{Kong2020}. 
In this work we only consider the acceleration of electrons in the perpendicular 
direction for the observations and simulations.

\subsection{Sample of SEPs with ICME-driven shocks from observations and
simulations}

\begin{figure}
\epsscale{1.}
\plotone{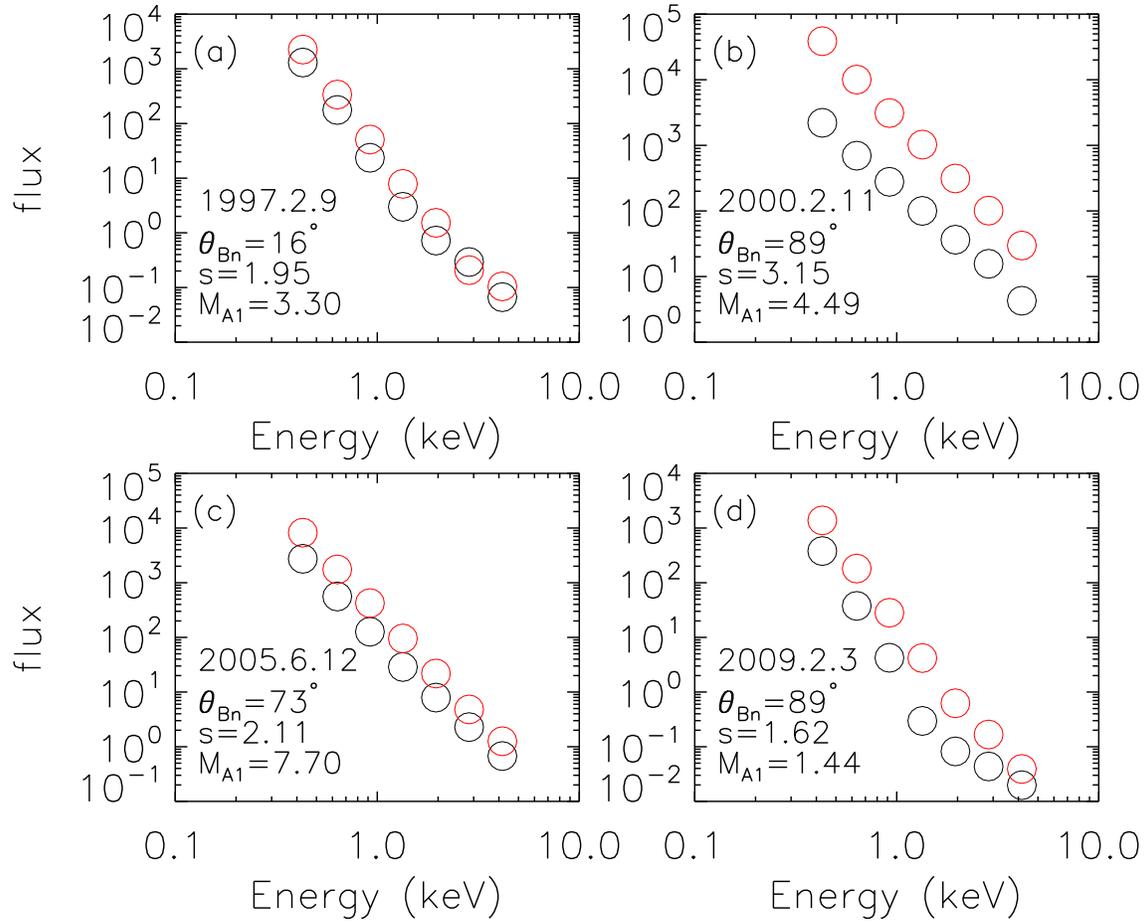}
        \figcaption{Downstream flux (red circles) and upstream flux (black
        circles) for four sample observed shock events in the energy
        range of 0.428 -- 4.161 keV. Note that the upstream and downstream flux
        are averaged over a period of 10 minutes. The date,
       shock angle, compression ratio, and Alfv$\acute{\text e}$n Mach
       number of each shock event are listed in the lower left corner.
\label{fig:fluxobser}}
\end{figure}

In order to show the shock acceleration we select four sample observed SEP
events from the 74 events with ICME-driven shocks listed in \citet{Yang2019}. In
Figure \ref{fig:fluxobser} red and black circles indicate downstream and
upstream fluxes in the $90^\circ$ pitch angle, respectively, for the four observed
shock events in the energy range of 0.428 -- 4.161 keV, averaged over a period
of 10 minutes. Each panel of the figure shows an event, with the date,
shock angle,
compression ratio, and Alfv$\acute{\text e}$n Mach number listed in the lower
left corner. The figure shows that for all the observed events the flux is 
larger in the downstream compared with the upstream, and there is significant
shock acceleration of energetic particles. 

\begin{figure}
\epsscale{1.}
\plotone{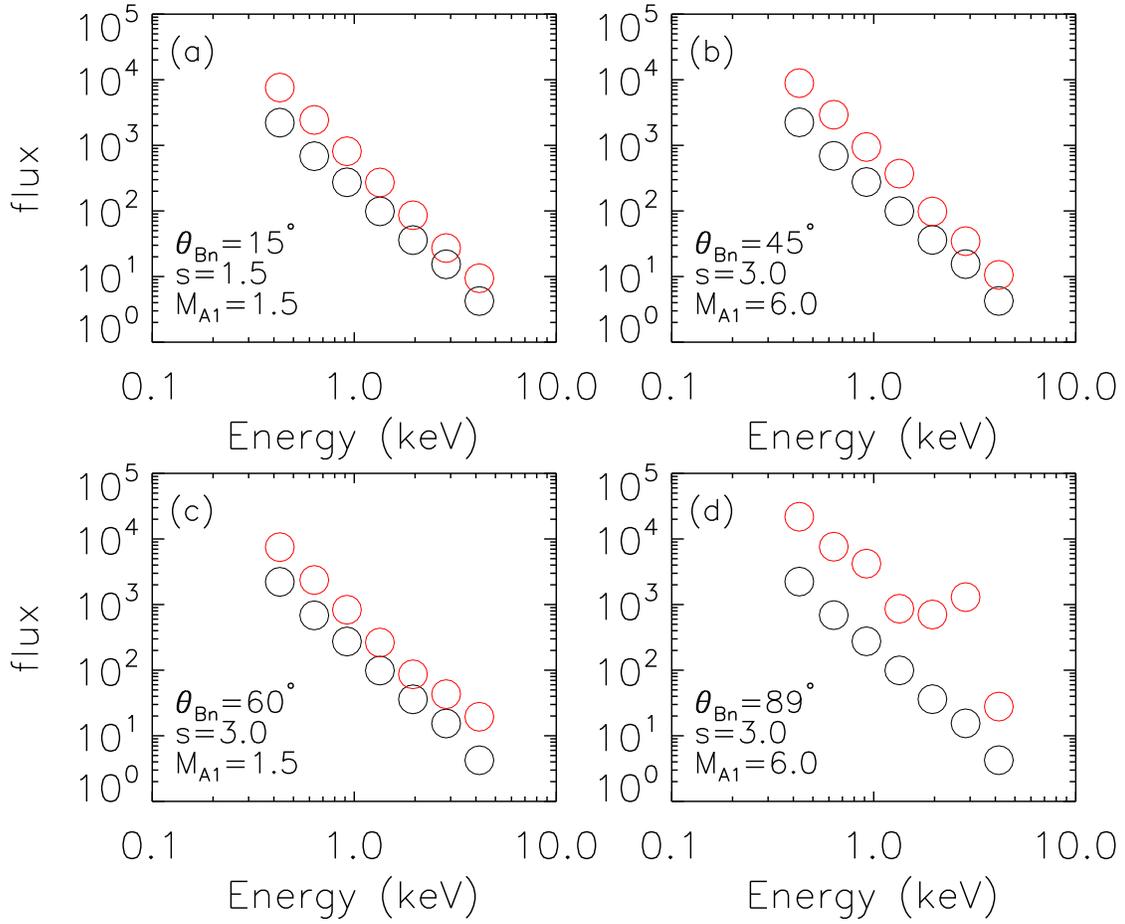}
        \figcaption{Downstream flux (red circles) 
        in the 90$^\circ$ pitch angle for four sample 
        simulation cases in the energy range of 0.428 -- 4.161 keV with
        the assumed source (black circles). 
        The shock angle, compression ratio,
        and Alfv$\acute{\text e}$n Mach number for each case are listed in the
        lower left corner. 
\label{fig:fluxsimu}}
\end{figure}

Furthermore, for each of the 315 simulation cases of the parameter combinations
 $(\theta_{\text{Bn}}, M_{\text{A}1}, s)$, we perform simulations with 10,000
 test particles using the backward-in-time method, with the 10-minute averaged 
 distribution in the $90^\circ$ pitch angle observed
on 2000 Feb 11 as the initial condition and obtain the downstream flux for 
 each of the energy channels $E_k$. The results of four selected sample 
 simulation cases are shown in Figure \ref{fig:fluxsimu}. The red circles
 in Figure \ref{fig:fluxsimu} indicate the average downstream flux 
 in the $90^\circ$ pitch angle from the simulations for the four cases in the 
 energy range of 0.428 -- 4.161 keV, and black circles show the initial distribution 
for simulations in the $90^\circ$ pitch angle. Similar as Figure
 \ref{fig:fluxobser}, each panel of Figure \ref{fig:fluxsimu} shows a simulation
 case, with the shock angle, compression ratio, and Alfv$\acute{\text
 e}$n Mach number listed in the lower left corner. It can seen that for all the
 four cases the simulated downstream flux is larger than the upstream flux which
 is used as the source for the simulations, and there is significant shock
 acceleration of particles in the simulations. 

These sample results of observations and simulations might suggests that
the shock acceleration efficiency increases with increasing shock angle
from parallel to perpendicular, and also that the acceleration efficiency 
increases with the increase of compression ratio and Alfv$\acute{\text e}$n 
Mach number for a similar shock angle.

\subsection{Shock acceleration efficiency from observations and simulations}

In order to further study the shock acceleration efficiency, for the 74
observational shock events we obtain the ratio, $R_a$, of downstream to
upstream flux at $90^\circ$ pitch angle in each of the energy channels $E_k$
listed in Table \ref{energychannels}. If the ratio $R_a$ is greater than a 
threshold value $R_t$, we suppose that the shock acceleration is efficient.
Here the value of $R_t$ is set to an arbitrary value of $4$. 
Actually,  we make a qualitative analysis, and the general
findings do not change if $R_t$ is set to other values, unless $R_t\sim 1$ or
$R_t\gg4$. We avoid to use a large $R_t$ in order to get good statistics. 
Furthermore, we set $R_t> 1$ to study the acceleration of electrons.
Therefore, we can study the effect of the shock angle 
$\theta_{\text{Bn}}$, upstream Alfv$\acute{\text e}$n Mach number
$M_{\text{A}1}$, and compression ratio $s$ on the shock
acceleration efficiency from the observations. First, we divide
the shock angle into different intervals, in each of which we obtain the
ratio $N_1/N_0$, where $N_1$ is the number of shock events with a high acceleration
efficiency and $N_0$ is the total number of shock events. Similarly, we 
divide the upstream Alfv$\acute{\text e}$n Mach number $M_{\text A1}$ and
compression ratio $s$ into different intervals, and we  obtain the ratio
$N_1/N_0$ in each interval of $M_{\text A1}$ and $s$, respectively. 

Furthermore, using the ratio, $R_a$, of the downstream flux from simulations
to the upstream flux we can determine that a shock with parameters
$(\theta_{\text{Bn}}, M_{\text{A}1}, s)$ could efficiently accelerate 
electrons if $R_a>R_t$ is satisfied. In this way, we can study 
the effect of $\theta_{\text{Bn}}$, $M_{\text{A}1}$, and $s$
on the shock acceleration efficiency from the simulations. 
The various values of shock parameters, such as $\theta_{\text{Bn}}$,
$M_{\text{A}1}$, and $s$ from the observations carry different weights.
However, for simplicity, we assume that the weight function is uniform.
The general findings of this work, in fact, will not change under
consideration of more realistic values of weight.
We divide the shock angle
into different intervals, in each of which we calculate the ratio $N_1/N_0$.
Similarly, we also divide $M_{\text{A}1}$ and $s$ into different intervals to
calculate the ratio $N_1/N_0$ in each interval of $M_{\text{A}1}$ and $s$, 
respectively.

\begin{figure}
\epsscale{1.}
\plotone{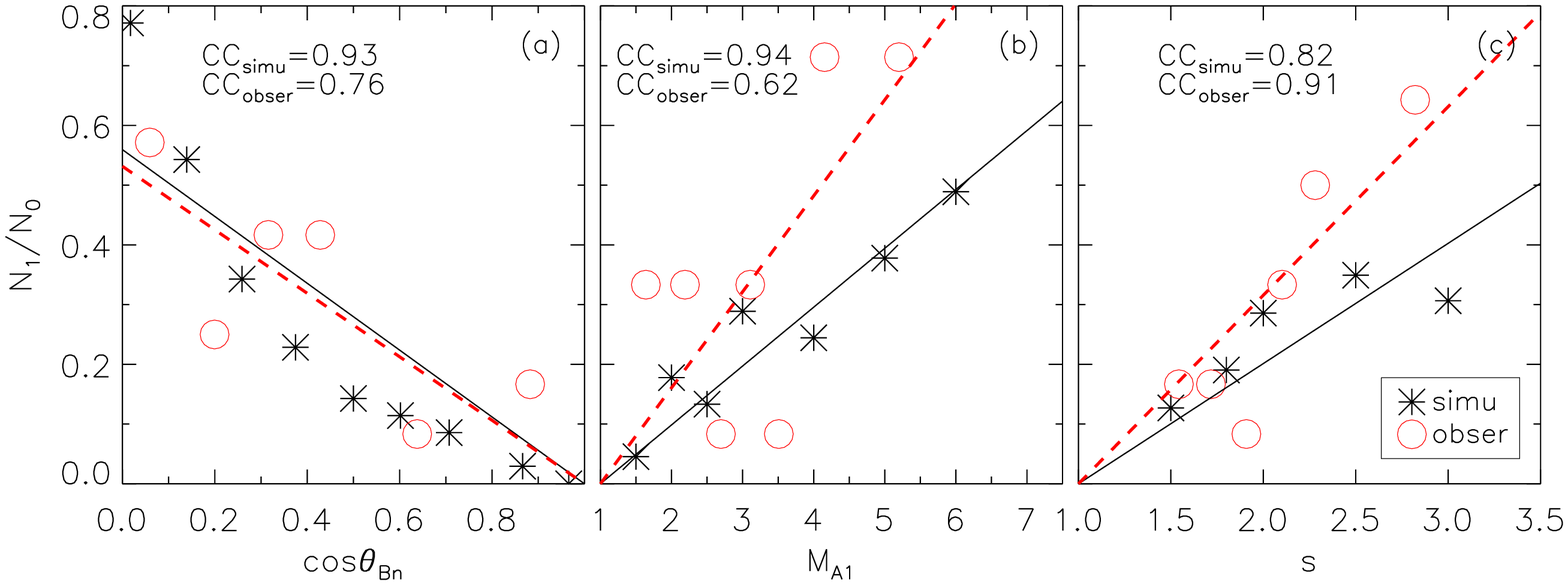}
        \figcaption{Ratio of event numbers $N_1/N_0$ ($N_1$ is the 
        number of shock events with a high acceleration efficiency 
        and $N_0$ is the total number of shock events in each interval), versus
the shock angle cosine $\cos\theta_{\text{Bn}}$ (left panel), 
upstream Alfv$\acute{\text e}$n Mach number $M_{\text A1}$ (middle panel), 
and compression ratio $s$ (right panel) in the energy channel of 0.428 keV for
the observations (circles) and simulations (asterisks), with the threshold 
value $R_t=4$ for the downstream to upstream flux ratio. Linear fits for the
observations and simulations are shown as red dashed and black solid lines, respectively. 
Note that $\text{CC}_{\text{simu}}$/$\text{CC}_{\text{obser}}$ denotes
the linear Pearson correlation coefficient of the y-axis values between the 
simulations/observations and fit line.
\label{fig:ratio0}}
\end{figure}

Figure \ref{fig:ratio0} shows the ratio $N_1/N_0$ versus the shock angle
cosine $\cos\theta_{\text{Bn}}$ (left panel), upstream 
Alfv$\acute{\text e}$n Mach number $M_{\text A1}$ (middle panel), and compression
ratio $s$ (right panel) in the energy channel of 0.428 keV for the
observations (circles) and simulations (asterisks). Note that we use $x_i$ with
$i=a$, $b$, and $c$ to indicate shock angle cosine 
$\cos\theta_{\text{Bn}}$, upstream Alfv$\acute{\text e}$n Mach number 
$M_{\text A1}$, and compression ratio $s$, respectively. 
Figure \ref{fig:ratio0}(a) shows that for both the observations and simulations,
with $\cos\theta_{\text{Bn}}\sim0$, i.e., quasi-perpendicular geometry, the
ratio $N_1/N_0$ is the highest, and it decreases with the increase of the
shock angle cosine $\cos\theta_{\text {Bn}}$. This indicates
that the acceleration of electrons is more efficient at quasi-perpendicular
shocks. In addition, it is shown that as $\cos\theta_{\text{Bn}}=1$, 
shock acceleration is very weak, i.e., the ratio $N_1/N_0\approx 0$.
Therefore, we assume the ratio $N_1/N_0$ for both the observations and
simulations can be fitted by a straight line
\begin{equation}
    y=\xi x-\xi,\label{equ:fitfunc}
\end{equation}
where $\xi$ is the fit parameter. Here, we fit data with $0\le x\le 1$. 
The red dashed and black solid lines in Figure \ref{fig:ratio0}(a) indicate the results 
obtained by fitting the data from the observations and simulations, respectively. 
The linear Pearson correlation coefficient of the y-axis values between the 
simulations/observations and fit line are given by $\text{CC}_{\text{simu}}$/$\text{CC}_{\text{obser}}$.
As shown in Figure \ref{fig:ratio0}(a), the slope $\xi$ is negative, and the good correlation 
indicates that the linear fit can roughly exhibit the trend of $N_1/N_0$ with the shock angle.
The fact that the value of $N_1/N_0$ increases with the decrease of 
$\cos\theta_{\text{Bn}}$, implies a stronger shock acceleration efficiency 
with increasing shock angle $\theta_{\text{Bn}}$. 

Figures \ref{fig:ratio0}(b) and (c) show that, if the
upstream Alfv$\acute{\text e}$n Mach number $M_{\text A1}=1$ and the
compression ratio $s=1$, respectively, the shock acceleration is very weak,
i.e., the ratio $N_1/N_0\approx 0$, for both the observations and simulations.
Due to the increase of ratio $N_1/N_0$ with increasing $M_{\text A1}$ and $s$, 
we assume that the variation of $N_1/N_0$ with $M_{\text A1}$ or $s$ for both 
the observations and simulations can be fitted by the straight line in Equation
(\ref{equ:fitfunc}) when $x\ge 1$. The red dashed and black solid lines indicate the
results obtained by fitting the data from the observations and simulations,
respectively. Also denoted in Figures \ref{fig:ratio0}(b)--(c) is the linear 
Pearson correlation coefficient for the simulations ($\text{CC}_{\text{simu}}$) 
and observations ($\text{CC}_{\text{obser}}$).
It is shown in Figures \ref{fig:ratio0}(b)--(c) that the slope 
$\xi$ is positive, and the fitting lines roughly exhibit the trend of $N_1/N_0$ 
with $M_{\text A1}$ and $s$. The increase of $N_1/N_0$ with $M_{\text A1}$ and $s$ 
indicates a stronger shock acceleration efficiency with
 the increase of $M_{\text A1}$ and $s$.

\begin{figure}
\epsscale{1.}
\plotone{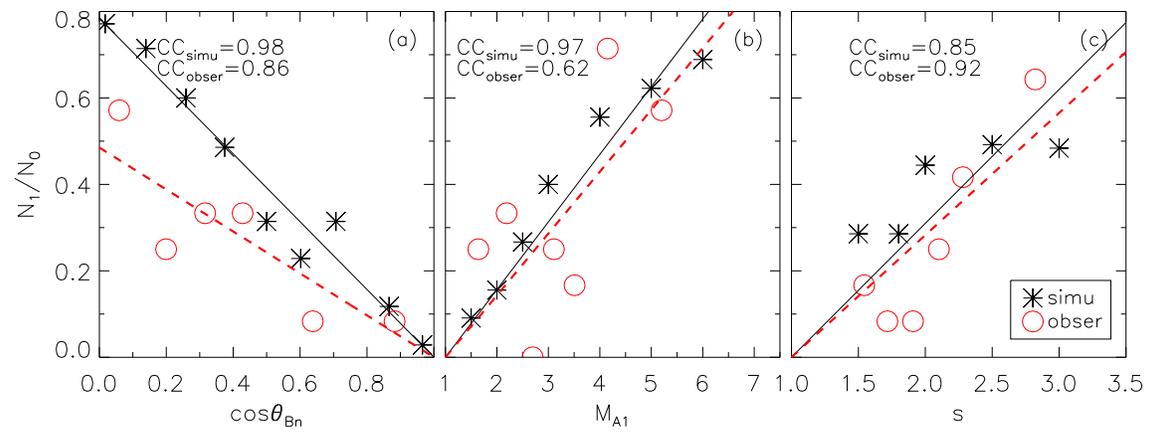}
        \figcaption{Same as Figure \ref{fig:ratio0}, except that the energy 
        channel is 0.634 keV.\label{fig:ratio1}}
\end{figure}

\begin{figure}
\epsscale{1.}
\plotone{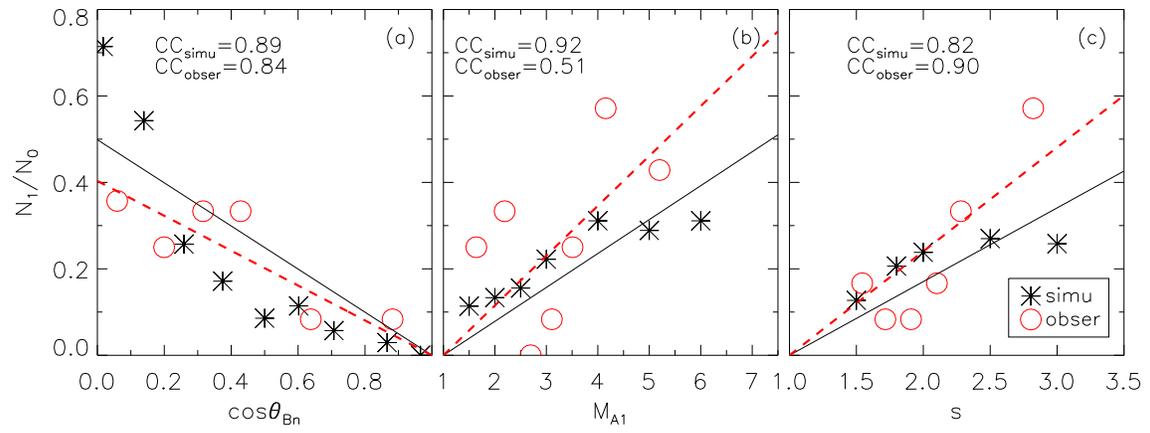}
        \figcaption{Same as Figure \ref{fig:ratio0}, except that the energy 
        channel is 0.920 keV.\label{fig:ratio2}}
\end{figure}

\begin{figure}
\epsscale{1.}
\plotone{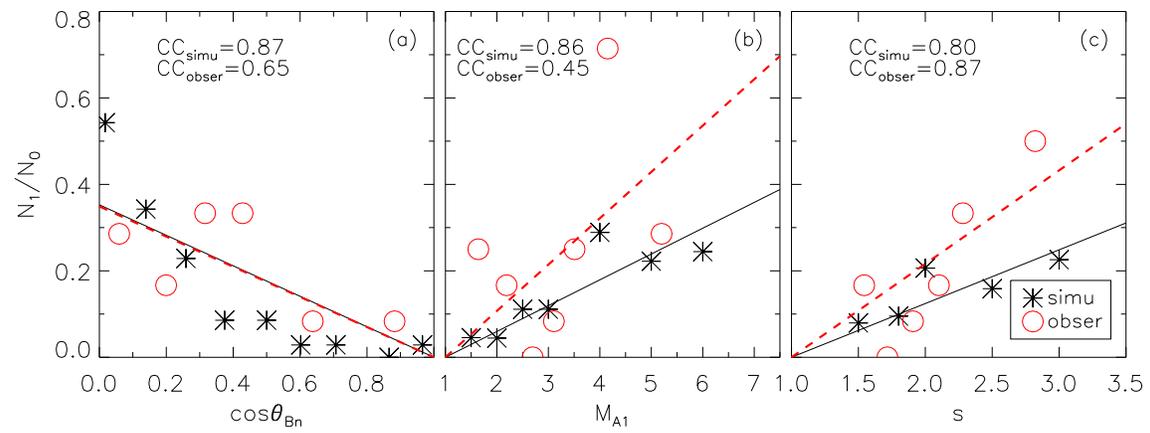}
        \figcaption{Same as Figure \ref{fig:ratio0}, except that the energy 
        channel is 1.339 keV.\label{fig:ratio3}}
\end{figure}

Figures \ref{fig:ratio1}, \ref{fig:ratio2}, and \ref{fig:ratio3} show
plots similar to Figure \ref{fig:ratio0}, except in the energy channels of
0.634, 0.920, and 1.339 keV, respectively. In these figures, both the 
observations (circles) and simulations (asterisks) show that the ratio of event
numbers with the high shock acceleration efficiency, $N_1/N_0$, decreases with
the shock angle cosine, and increases with the upstream
Alfv$\acute{\text e}$n Mach number and compression ratio.  For other energy
channels listed in Table \ref{energychannels} we have similar results.
It is noted that the results from observations and simulations from  
Figures \ref{fig:ratio0}, \ref{fig:ratio1}, \ref{fig:ratio2}, and
\ref{fig:ratio3} show quite large scatterings. The reason may be the limit of 
numbers $N_0$ and $N_1$ from both observations and simulations.

\begin{figure}
\epsscale{.5}
\plotone{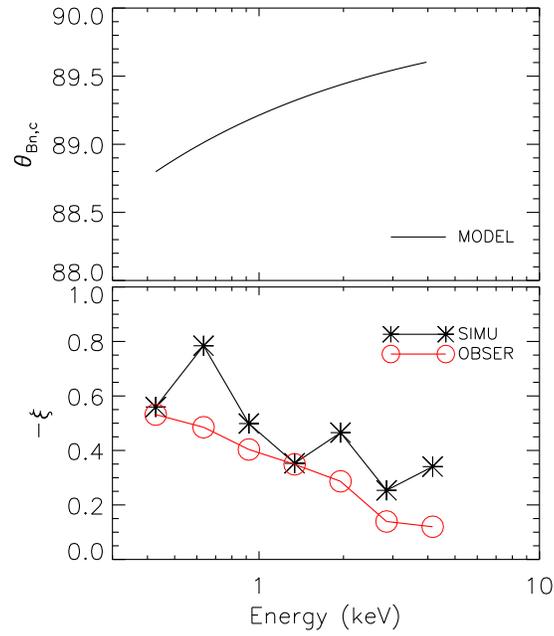}
        \figcaption{Top panel: modeling results of the 
        critical shock angle $\theta_{\text{Bn,c}}$ 
        (Equation (\ref{equ:theta_b})) versus electron energy.
        Bottom panel: negative slope, $-\xi$, in each energy channel
        by a linear fit of $N_1/N_0$--$\cos\theta_{\text{Bn}}$ to
Equation (\ref{equ:fitfunc})  
        for the simulations (black asterisks) and observations (red
        circles).
\label{fig:theta_xi}}
\end{figure}

We assume that the shock acceleration efficiency is prominent when the
shock angle $\theta_{\text{Bn}}$ is approximately or greater than the critical
shock angle $\theta_{\text{Bn,c}}$. The ratio
$R_{\theta}\equiv\theta_{\text{Bn}}/\theta_{\text{Bn,c}}$ can be used to show
the shock acceleration efficiency. If the critical shock angle is smaller, the
ratio $R_{\theta}$ would become larger, so that the electron shock acceleration 
is more efficient. In the top panel of Figure \ref{fig:theta_xi} we show the
modeling results of the critical shock angle $\theta_{\text{Bn,c}}$ from Equation
(\ref{equ:theta_b}). It is seen that the critical shock angle increases 
with the increase of electron energy, and 
is very close to $90^\circ$ for the highest energy channel. The bottom panel of
Figure \ref{fig:theta_xi} shows the negative slope, $-\xi$, from the linear fit
of $N_1/N_0$--$\cos\theta_{\text{Bn}}$ in each energy channel as shown above for
both the simulations (black asterisks) and observations (red circles). We can 
see that $-\xi$ from the data of observations roughly agrees with the result from 
simulations. Moreover, the value of $-\xi$ generally decreases with increasing 
electron energy, 
i.e., the shock acceleration is less efficient with the increase of electron energy. 
In other words, this suggests that higher energy electrons are more difficult to be
accelerated by shocks.

\begin{figure}
\epsscale{.5}
\plotone{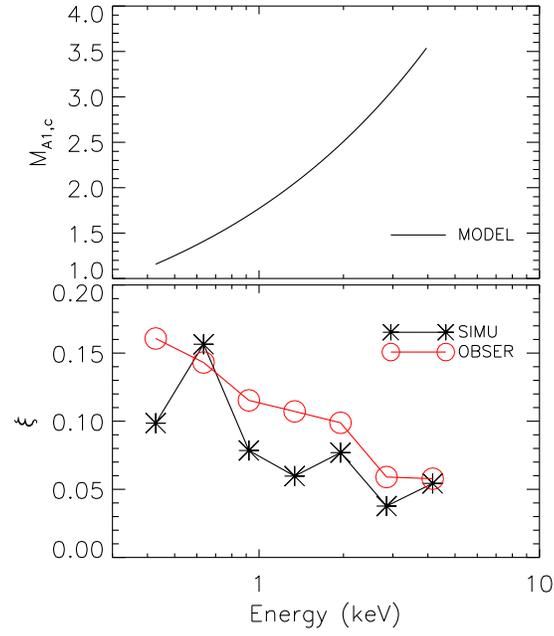}
        \figcaption{
       Top panel: modeling results of the 
        critical upstream Alfv$\acute{\text e}$n Mach number $M_{\text{A1,c}}$ 
        (Equation (\ref{equ:M_A1b})) versus electron energy.
        Bottom panel: slope, $\xi$, in each energy channel
        by a linear fit of $N_1/N_0$--$M_{\text{A1}}$ to
Equation (\ref{equ:fitfunc})  
        for the simulations (black asterisks) and observations (red
        circles).
\label{fig:Ma1_xi}}
\end{figure}

\begin{figure}
\epsscale{.5}
\plotone{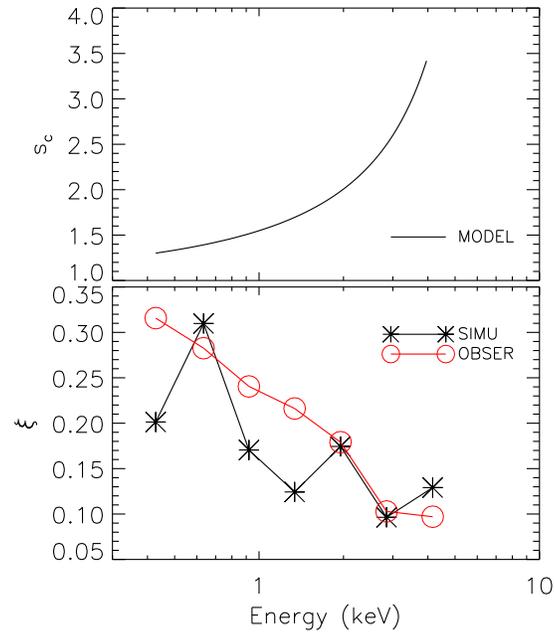}
        \figcaption{
       Top panel: modeling results of the 
        critical compress ratio $s_{\text{c}}$ 
        (Equation (\ref{equ:s_b})) versus electron energy.
        Bottom panel: slope, $\xi$, in each energy channel
        by a linear fit of $N_1/N_0$--$s$ to
Equation (\ref{equ:fitfunc}) 
        for the simulations (black asterisks) and observations (red
        circles).
\label{fig:s_xi}}
\end{figure}

Suppose the shock acceleration is strong when the upstream 
Alfv$\acute{\text e}$n Mach number $M_{\text{A1}}$ and the compression ratio
$s$ are approximately or greater than their critical values, $M_{\text{A1,c}}$
and $s_{\text c}$, respectively. The ratios $R_M\equiv M_{\text{A1}}/M_{\text{A1,c}}$ 
and $R_s\equiv s/s_{\text c}$ can be used to indicate the efficiency of shock acceleration. 
If $M_{\text{A1,c}}$ and $s_{\text c}$ are smaller, $R_M$ and $R_s$, respectively, 
would be greater, so that the electron shock acceleration is more efficient. 
In the top panels of Figures \ref{fig:Ma1_xi} and \ref{fig:s_xi}
we show the modeling results from Equations (\ref{equ:M_A1b}) and
(\ref{equ:s_b}) for the critical upstream Alfv$\acute{\text e}$n Mach number
$M_{\text{A1,c}}$ and the critical compression ratio $s_{\text{c}}$,
respectively, with the efficient shock acceleration versus electron energy.
Note that the modeling results for $M_{\text{A1,c}}$ and $s_{\text c}$ depend
on the characteristic parameters $s_{\text{r}}$ and $M_{\text{r}}$, 
respectively. Because these modeling results are only used to qualitatively
show the trends of the shock acceleration efficiency, here we arbitrarily set
$s_{\text{r}}=2$ and $M_{\text{r}}=2.5$. The modeling results show that
$M_{\text{A1,c}}$ and $s_{\text c}$ increase with the increase of electron
energy. Therefore, the shock acceleration is less efficient for higher
energy electrons. The bottom panels of Figures \ref{fig:Ma1_xi} and \ref{fig:s_xi}
show the slope, $\xi$, from the linear fit of $N_1/N_0$--$M_{\text{A1}}$ and
$N_1/N_0$--$s$, respectively, in each energy channel as shown above, for both
the simulations (black asterisks) and observations (red circles). It is shown
that the slope $\xi$ generally decreases with electron energy, i.e., the shock
acceleration is less efficient with the increase of electron energy. Moreover, 
the slope from the data of observations is in rough agreement with the result 
from the simulations.
From Figures \ref{fig:theta_xi}-\ref{fig:s_xi} it is seen that the
curves from simulations are less smooth than those from observations. 
The reason may be that the physical models we used have some shortcomings, and 
the computational resources are limited, so that our simulation results are 
not sufficiently accurate to fully reproduce the observations.
\section{summary and discussion}
\label{sec:summary}

We study the shock acceleration efficiency of suprathermal electrons, using
spacecraft observational data for some ICME-driven shock events during 
1995--2014 and test-particle simulations for cases of different shock
parameters. It is suggested by \citet{Yang2018} \citep[see also,][]{Kong2020}
that there exists a stream of energetic particles anti-sunward downstream of 
the shock in some SEP events. We check the 74 ICME-driven shock events listed 
in \citet{Yang2019}, and find that for 65 events with complete 
data there is an anti-sunward beam of electrons downstream of the shock. 
This downstream anti-sunward beam may originate from the background 
solar wind rather than the shock acceleration. We need to use
the sunward particles to study energetic
particles accelerated by the shock along the magnetic field direction. It is
shown that for most observational cases the acceleration of electrons in the 
sunward direction is much less efficient than in the perpendicular direction. 
Consequently, we suggest that the strongest acceleration occurs in the 
perpendicular pitch-angle direction. Therefore, in this paper the shock 
acceleration efficiency of electrons at $90^\circ$ pitch angle is 
investigated by comparing the simulations with observations.

For the 74 shock events listed in \citet{Yang2019}, using data analysis
on the spacecraft observations in seven energy channels ranging from 0.428 keV 
to 4.161 keV, we study how the efficiency of shock acceleration depends on
different shock parameters, including the shock angle $\theta_{\text{Bn}}$,
upstream Alfv$\acute{\text e}$n Mach number $M_{\text A1}$, and compression 
ratio $s$. In addition, we perform a similar research using data from 315
test-particle simulation cases with different shock parameters. With
data from both the observations and simulations, we obtain the ratio of
the average downstream to upstream flux $R_a$ for each shock acceleration case.
Next, we get the ratio $N_1/N_0$ of the events with a large $R_a$, i.e., $R_a>R_t$
(the threshold value $R_t$ set to $4$ here). 
We find that for both the observations and simulations, 
the ratio $N_1/N_0$ decreases with the increase of shock angle
cosine and increases with upstream Mach number and compression ratio. This 
indicates that a large shock angle, Mach number, and compression ratio
can contribute to enhance the shock acceleration efficiency. 
Furthermore, the variation of ratio $N_1/N_0$ with respect to the shock angle cosine
$\cos\theta_{\text{Bn}}$, upstream Alfv$\acute{\text e}$n Mach number 
$M_{\text A1}$, and compression ratio $s$ is fitted by a straight line of
slope $\xi$ through the point $(1,0)$ for the observations and simulations in
different energy channels. The value of slope $\xi$ by fitting  
$N_1/N_0$ --$\cos\theta_{\text{Bn}}$, --$M_{\text A1}$, and --$s$, respectively,
shows that the shock acceleration is more efficient for lower energy particles
compared to higher energy particles from the observations and simulations,
which is obtained over a relatively small energy range. The reason for 
the lower acceleration efficiency in higher energy may be the following.
\textbf{With stronger scattering of particles by turbulence, the period
of each cycle of the particle crossing of the shock front is shorter and the
acceleration is more efficient \citep{Drury1983,Kong2019}. Furthermore,}
the turbulence scatters lower energy particles more easily compared to
higher energy particles, so low-energy particles are accelerated
easily to high energies by the shock. In addition, lower energy particles are
more likely to stay in the shock transition  within drift time. On the other hand, 
particles accelerated to higher energies may be firstly accelerated to lower
energies.
It is noted that
if $R_t$ is set to some other values, the ratio $N_1/N_0$ probably vary, but 
the general findings of this paper will be the same.

We develop a model for the value of critical shock angle
$\theta_{\text{Bn,c}}$ with high efficiency of shock acceleration, assuming
shock drift acceleration of particles. We also obtain models for the critical 
values of upstream Mach number $M_{\text{A1,c}}$ and compression ratio
$s_{\text{c}}$ with efficient acceleration, based on the average momentum change
of particles proposed by \citet{Drury1983} for each cycle of the particle
crossing of the shock front. Although there are undetermined parameters, the
models may be used to qualitatively show the trend of the shock acceleration of
particles. It is assumed that the shock acceleration efficiency is prominent 
when the shock angle, upstream Alfv$\acute{\text e}$n Mach number, and compression ratio are
larger than their respective critical values. Consequently, if the
critical values of shock angle, upstream Alfv$\acute{\text e}$n Mach number, 
and compression ratio are smaller, the electron shock acceleration is more
efficient. The modeling results show that the critical values of shock angle,
upstream Alfv$\acute{\text e}$n Mach number, and compression ratio 
increase with the increase of particle energy, implying stronger
shock acceleration for lower energy particles.
Therefore, the data of observations and simulations, as well as 
the theoretical modeling results, suggest that shock drift acceleration
is efficient in the electron acceleration by ICME-driven shocks 
\citep{Yang2018, Yang2019, Kong2020}.

It should be noted that,  
the upper limit of slab turbulence wavenumbers in the simulations is set 
to $k_M\sim1.75\times10^{-4}$ m$^{-1}$. It is, however, not
significantly larger
than the minimum resonant wavenumber ($k_{\text{rm}}=eB_0/p_{\text e}$, where
$p_{\text e}$ is the electron momentum) of low energy electrons. For example, 
for electrons with an energy of 0.428 and 0.634 keV, the minimum resonant
wavenumber is $k_{\text{rm}}\sim 1\times 10^{-4}$ m$^{-1}$ and 
$8\times 10^{-5}$ m$^{-1}$, respectively. The 
break wavenumber $k_{\text b}$ for the slab turbulence between the inertial and 
dissipation ranges is $10^{-6}~\text{m}^{-1}$. Therefore for electrons in
the energy range of $0.428$ keV to $4.161$ keV the resonant wavenumber locates in 
the dissipation range of the slab turbulence.
In addition, the upper limit of the wavenumber of 2D magnetic
turbulence component is $6.07\times 10^{-7}~\text m^{-1}$, 
and the dissipation range of 2D turbulence is ignored.
It is considered that energetic particles are in resonance with slab turbulence 
in the resonant wavenumber $k_r$, and parallel diffusion depends on the slab
component with the assumption of quasi-linear theory, so the upper limit of the
wavenumber of slab component of the magnetic turbulence is very important. We
also consider that perpendicular diffusion depends on both the 2D and slab
turbulence by the non-linear effects. It is more difficult to 
construct a 2D component turbulence with a very large wavenumber range.
In addition, lower wavenumber range of 2D component can influence the
perpendicular diffusion because of the non-linear effects. Therefore, we set a
2D component with a low upper limit of wavenumber.

As a result, in our simulations lower 
energy electrons have a narrow resonant range in the power spectrum of the turbulence
compared to the observations. 
The simulations of the acceleration of high energy electrons, which
have a large velocity, require more computational resources in order to 
ensure numerical accuracy. Therefore, 
we expect to use more powerful computational resources to increase 
the upper limit of the slab turbulence and the computing accuracy in
the simulations in future works.

\acknowledgments
This work was supported, in part, under grants NNSFC 42074206 and NNSFC
41874206. 
The work was carried out at National Supercomputer Center 
in Tianjin, and the calculations were performed on TianHe-1 (A).


\clearpage

\clearpage

\clearpage

\clearpage

\clearpage

\clearpage

\clearpage

\clearpage

\clearpage

\clearpage

\clearpage

\clearpage

\clearpage
\begin{table}
\begin{center}
\caption {Energy Channels $E_k$\label{energychannels}}
\begin{tabular} {l|lllllll}\hline\hline
$k$ & 1 & 2 & 3 &4 & 5 & 6 & 7\\ \hline
$E_k$ (keV)  & 0.428 & 0.634 & 0.920 & 1.339 & 1.952 & 2.849 & 4.161 \\\hline
\end{tabular}
\end{center}
\end{table}

\clearpage
\begin{table}
\begin{center}
\caption {Parameter Values for Various Simulation Cases\label{casespara}}
\begin{tabular} {l|l}\hline\hline
Parameter & Values \\\hline
$\theta_{\text{Bn}}$ & $15^\circ$, $30^\circ$, $45^\circ$, $53^\circ$,
$60^\circ$, $68^\circ$, $75^\circ$, $82^\circ$, $89^\circ$\\
$M_{\text{A}1}$ & 1.5, 2.0, 2.5, 3.0, 4.0, 5.0, 6.0\\
$s$ & 1.5, 1.8, 2.0, 2.5, 3.0\\\hline
\end{tabular}
\end{center}
\end{table}

\clearpage
\begin{table}
\begin{center}
\caption {Input Parameters for the Shock and Turbulence\label{shockturbpara}}
\begin{tabular} {l|l|l}\hline\hline
Parameter & Description & Value \\\hline
$B_{01}$   & upstream magnetic field &  7.0 nT \\
$L_{\text{th}}$ & shock thickness & 2$\times10^{-6}$ au \\
$V_{\text{A1}}$ & upstream Alfv$\acute{\text e}$n speed & 
$67~\text{km}~\text s^{-1}$\\
$\lambda$  & slab correlation length & 0.02 au   \\
$\lambda_x$  & 2D correlation length & $\lambda/2.6$   \\
$E_{\text{slab}}:E_{\text{2D}}$ &two-component energy density ratio &
$0.25$\\
${\left(b/B_0\right)^2}_1$ & upstream turbulence level & 0.25    \\
${\left(b/B_0\right)^2}_2$ & downstream turbulence level & 0.36    \\
$k_{\text b}$ & break wavenumber of slab turbulence& 10$^{-6}$ m$^{-1}$ \\
$\beta_{\text i}$ & inertial spectral index & 5/3\\
$\beta_{\text d}$ & dissipation spectral index & 2.7\\\hline
\end{tabular}
\end{center}
\end{table}

\end{document}